# Magneto-dielectric and Magneto-resistive in the Mixed Spinel – MgFe$_2$O$_4$

B. Santhosh Kumar[1], N. Praveen Shankar[1], N. Aparna devi[1,2] and C.Venkateswaran[1]*

1. Department of Nuclear Physics, University of Madras, Guindy campus, Chennai – 600 025, India.
2. Department of Physics, Ethiraj College for Women, Egmore, Chennai, India – 600 008.
*Corresponding author E mail: cvunom@hotmail.com

**Abstract.** The mixed spinel, MgFe$_2$O$_4$ has been synthesized by ball-milling assisted sintering method. X-ray diffraction study confirms formation of cubic MgFe$_2$O$_4$ and the lattice parameter values calculated are $a = b = c = 8.369(3)$ Å. Vibrating sample magnetometer measurements at room temperature shows a soft ferrimagnetic nature. Magneto-Dielectric and Magneto-Restive plots confirm coupling at room temperature in the prepared MgFe$_2$O$_4$. The peak at 500 Oe in the MD plot is due to the canting of Fe$^{3+}$ ions distributed in octahedral and tetrahedral sites.

## INTRODUCTION

Spinel ferrites of the AFe$_2$O$_4$ type ( A = Mn, Co, Ni, Mg and Zn) are an important class of magnetic materials because of their applications in magnetic sensors, magnetic core, high speed digital tape or disk recording[1]. Ferrites usually have high chemical and thermal stability with a reproducible physical property. Magnesium ferrites are one among this class of ferrites due to (i) distribution of Fe$^{3+}$ ions in octa and tetrahedral voids, (ii) slow saturation magnetization value which makes them suitable for memory switching purposes[2].

Magnesium (MgFe$_2$O$_4$) ferrite, also called ferrospinels, crystallizes in the cubic structure with *Fd3m* space group. A formula unit of MgFe$_2$O$_4$ totally has 56 ions per unit cell and the void around oxygen gives rise two to tetrahedral and octahedral voids leading to many interesting properties[2]. If the A sites are completely occupied by divalent ions and the B site by trivalent ions then it is known as spinel ferrite. In the case of inverse spinel, a divalent metal occupies B site and a trivalent ion occupies A site. Mixed spinel's have the divalent and trivalent metal ions distributed among A and B sites [2]. MgFe$_2$O$_4$ is an important mixed spinel compound where the distribution of metal ions can be written as {(Mg$_{0.1}$Fe$_{0.9}$)$_{A\text{-site}}$ (Mg$_{0.9}$Fe$_{1.1}$)$_{B\text{-site}}$O$_4$} in which 0.1 % of Mg$^{2+}$ ions are in A-site and the rest in B-site[3].

## Experimental

MgO and Fe$_2$O$_3$ in stoichiometric ratio were taken in zirconia vial with a ball to powder ratio of 27:10, milled at 250 rpm for 5 hour and sintered at 1200 °C for 5 hours in a box furnace. The prepared ferrite sample is analysed by X-ray diffraction (XRD) and vibrating sample magnetometer (EG & G PARC 4500) measurements. Magneto-dielectric (MD) and magneto-resistive (MR) properties at room temperature are studied using field dependent impedance spectroscopy (Solatron 1260) in the frequency range of 10 KHz to 10 Hz[4].

## Results and Discussions

Fig 1 shows the XRD pattern of the prepared sample which clearly shows the formation of phase pure cubic MgFe$_2$O$_4$ with space group *Fd3m*. The cell parameters calculated are $a = b = c = 8.369(3)$ Å which agrees with the JCPDS file no: 89-4924[5,6].

Vibrating sample magnetometer measurements confirm the soft ferrimagnetic nature (Fig 2.). The saturation magnetisation is 22.3 emu/g (0.91 $\mu_B$), retentivity is 2 emu/g, with a small coercivity. The saturation magnetisation of MgFe$_2$O$_4$ depends on the distribution of Fe$^{3+}$ ions, and varies with synthesis methodology. In an ideal mixed spinel, 0.9 of Fe$^{3+}$ in A-site and 1.1 Fe$^{3+}$ in B-site are aligned opposite to each other. The resultant net magnetic moment of 0.2 Fe$^{3+}$ (each Fe$^{3+}$ is associated with 5 $\mu_B$ and Mg$^{2+}$ has 0 $\mu_B$) corresponds to 1 $\mu_B$. But VSM confirms 0.91$\mu_B$, which concludes the Fe$^{3+}$ distribution as 0.91 % at A and 1.09 % at B. In the absence of oxygen vacancy, the net magnetic moment formulae for the prepared sample can be written as {(Mg$_{0.09}$Fe$_{0.91}$) $_{A-site}$ (Mg$_{0.91}$Fe$_{1.09}$) $_{B-site}$O$_4$} which results in (1.09-0.91=0.18~0.91µ$_B$).

increase in frequency due to the dominance of Fe ions at higher frequency.

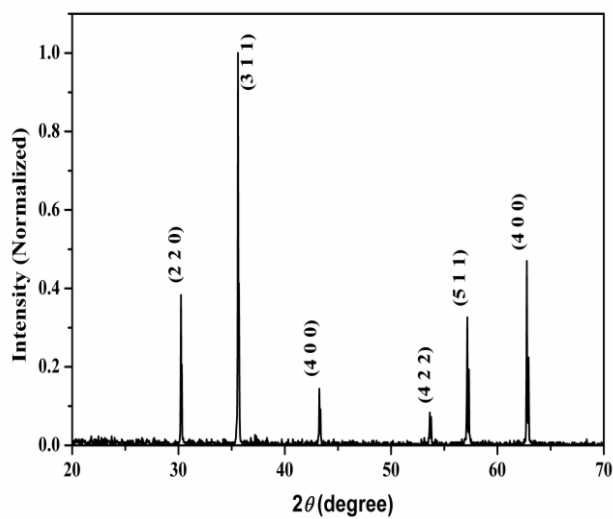

**Figure 1.** RT XRD pattern of MgFe$_2$O$_4$

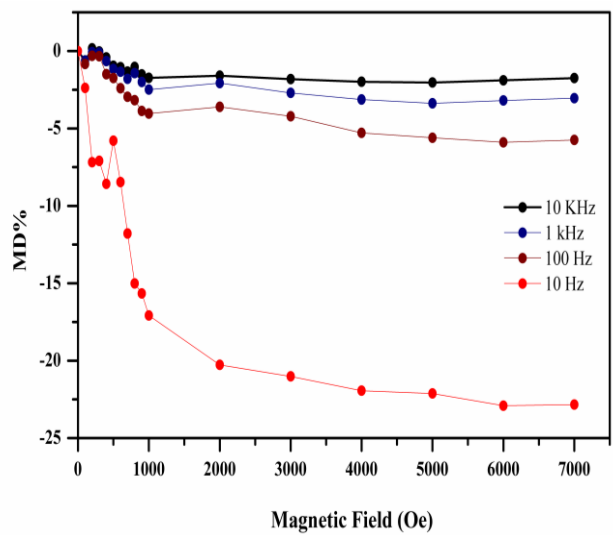

**Figure 3.** Magneto-dielectric coupling of MgFe$_2$O$_4$ at various frequency

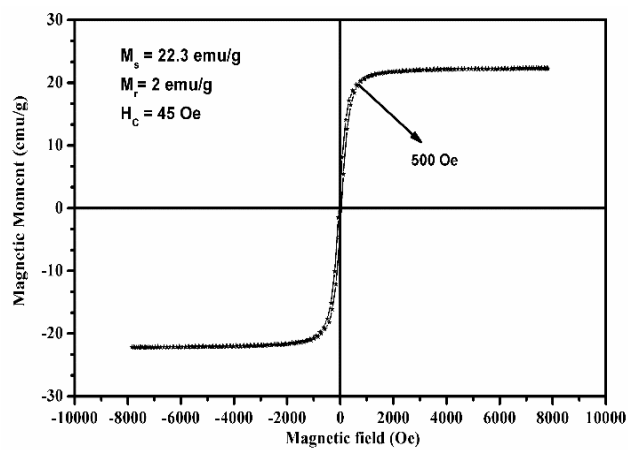

**Figure 2.** Hysteresis plot of mixed spinel MgFe$_2$O$_4$

The magneto-dielectric coupling has been studied using impedance spectroscopy by subjecting the sample to a frequency range of 10 KHz to 10 Hz under external magnetic field (0 Oe to 7000 Oe). The MD% = {(ε(H)-ε(0))/ ε(0))X100},in Fig 3, shows that the change in MD coupling reaches a maximum of 25% at 10 Hz at an applied magnetic field of 7000 Oe[4,7]. This could be due to the response of Fe$^{3+}$ ion with respect to the external magnetic field. The peak at 500 Oe due to the alignment of Fe ions also correlates with the VSM plot at 500 Oe where the magnetisation starts saturating. The MD response decreases with

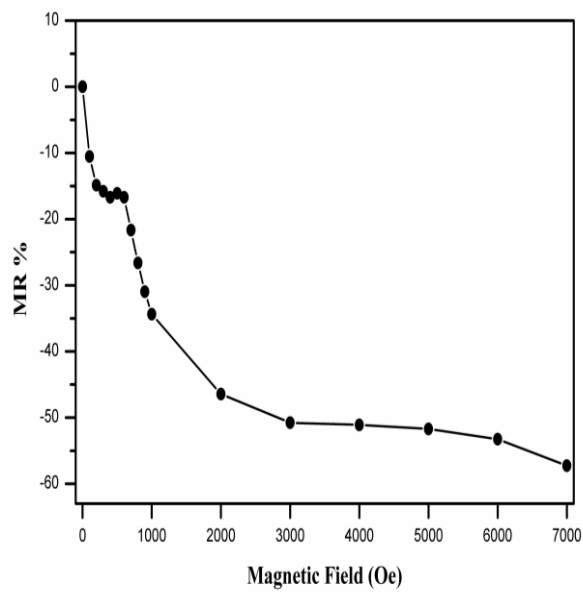

**Figure 4.** Field dependent magento-resistive coupling

The magneto-resistive (MR) plot as a function of the applied field (Fig 4) shows direct coupling and reaches 50% at 7000 Oe, which is twofold that of the MD response. Absence of peak at 500 Oe unlike MD coupling confirms the absence of Fe$^{3+}$ ion contribution in resistance.

## Conclusion

$MgFe_2O_4$ mixed spinel synthesised by ball milling and subsequent sintering at 1200 °C shows soft ferrimagnetic nature at room temperature. The $M_s$ Value of 22.3 emu/g (0.91 $\mu_B$) corresponds to the spinel formula unit {$(Mg_{0.09}Fe_{0.91})_{A-site} (Mg_{0.91}Fe_{1.09})_{B-site}O_4$}. Field dependent impedance spectroscopy confirms magneto-dielectric and magneto-resistive coupling at room temperature. The MD coupling is found to 25% at 7000 Oe at 10 Hz and in the higher frequency regime the coupling dominates. MR is found to be 50% at 7000 Oe which is due to the distribution of $Fe^{3+}$ in the A and B sites of spinel.

## ACKNOWLEDGMENTS

The author B. Santhosh kumar (BSK) thank DST – Inspire for its financial support in the form of fellowship and also Mr B. Soundararajan, technical officer for his kind help and support.